\documentclass[a4paper,11pt]{article}
\usepackage{jinstpub} 
\usepackage{lineno}
\usepackage{orcidlink}
\usepackage{graphicx}
\hypersetup{pageanchor=false}

\title{Installation and performance of the 3rd Veto plane at the SND@LHC detector}

\collaboration{The SND@LHC Collaboration}
\author[9]{D.~Abbaneo~\orcidlink{0000-0001-9416-1742}}
\author[1,2]{G.~Acampora~\orcidlink{0000-0003-4082-5616}}
\author[42]{S.~Ahmad~\orcidlink{0000-0001-8236-6134}}
\author[1,2]{R.~Albanese~\orcidlink{0000-0003-4586-8068}}
\author[1]{A.~Alexandrov~\orcidlink{0000-0002-1813-1485}}
\author[1,2]{F.~Alicante~\orcidlink{0009-0003-3240-830X}}
\author[1,2]{F.~Aloschi~\orcidlink{0000-0002-2501-7525}}
\author[6]{K.~Androsov~\orcidlink{0000-0003-2694-6542}}
\author[3]{A.~Anokhina~\orcidlink{0000-0002-4654-4535}}
\author[38]{C.~Asawatangtrakuldee~\orcidlink{0000-0003-2234-7219}}
% ...

\affiliation[1]{Sezione INFN di Napoli, Napoli, 80126, Italy}
\affiliation[2]{Universit\`{a} di Napoli ``Federico II'', Napoli, 80126, Italy}
\affiliation[3]{Affiliated with an institute formerly covered by a cooperation agreement with CERN}
% ...
\affiliation[6]{Institute of Physics, EPFL, Lausanne, 1015, Switzerland}
% ...
\affiliation[9]{European Organization for Nuclear Research (CERN), Geneva, 1211, Switzerland}
% ...
\affiliation[38]{Chulalongkorn University, Bangkok, 10330, Thailand}
% ...
\affiliation[42]{Affiliated with Pakistan Institute of Nuclear Science and Technology (PINSTECH), Nilore, 45650, Islamabad, Pakistan}
% ...

\emailAdd{giulia.paggi@cern.ch}

\abstract{ 
During 2022/2023 the optimal inefficiency of the Veto system of the SND@LHC detector was measured to be $(7.8\pm2.8)\times10^{-8}$. 
To reduce this inefficiency, a third Veto plane was installed during the 2023-2024 Year End Technical Stop. In addition, the Veto system was lowered to cover the target fully, thereby increasing acceptance. 
This paper describes how the inefficiency of the Veto system was reduced from $7.8\times10^{-8}$ with an acceptance of about $64\%$ of the target area in 2022-2023 to $(4.9\pm1.9)\times10^{-9}$ on the full area in 2024.}


\begin{document}
\maketitle
\flushbottom

\section{Introduction}
\label{sec:intro}
\begin{figure}[ht]
\centering
\includegraphics[width=\textwidth]{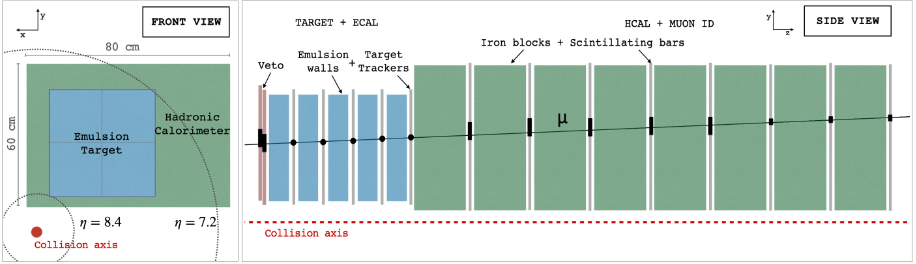}
\caption{Layout of the SND@LHC detector} 
\label{fig:detector}
\end{figure}

The Scattering and Neutrino Detector at the LHC (SND@LHC) is a compact, stand-alone experiment to perform measurements with neutrinos produced at the LHC in the pseudo-rapidity region of ${{7.2} < \eta < {8.4}}$ \cite{SNDLHCdetectorpaper}. The experiment is located 480 m downstream of the ATLAS interaction point, in the unused TI18 tunnel. It comprises a Veto system to identify incoming charged particles (mostly muons), a target region where neutrinos interact, a hadronic calorimeter (HCAL) and a muon identification system. The target region consists of bricks of emulsion cloud chambers. The emulsion cloud chambers bricks are interleaved with scintillating fibre (SciFi) tracker layers, which allow the identification of candidates for neutrino interactions. The Veto, HCAL and Muon systems use scintillating bars with different geometries:  6 cm bars in the Veto and upstream part of the HCAL, to achieve a high detection efficiency and energy resolution respectively, and 1 cm bars in the downstream part, to effectively isolate and measure the position and angle of crossing muons. In 2022, the Veto system had two Veto planes. One Veto plane consists of 7 bars and each bar is read by 16
silicon photomultipliers (SiPMs), 8 on each end (see Figure~\ref{fig:detector}).

Using data collected in 2022, we reported the observation of neutrino interactions from $\nu_{\mu}$s originating at the LHC IP1~\cite{Albanese:031802}. To obtain this result it was necessary to achieve a pure set of neutrino signals over background by applying selection cuts with a strong rejection power. 

\begin{figure}[ht]
\centering
\includegraphics[width=0.5\textwidth]{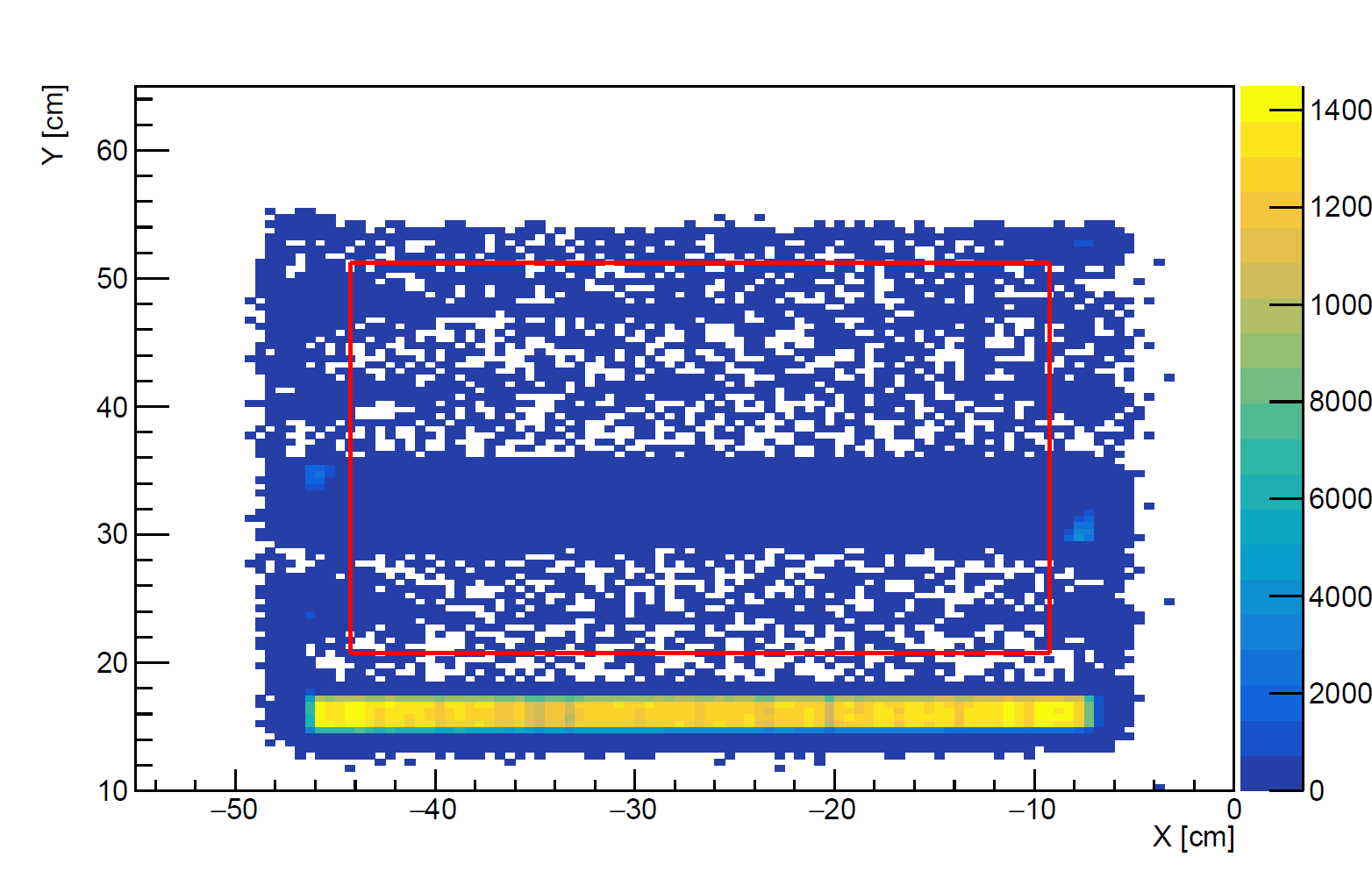}\includegraphics[width=0.5\textwidth]{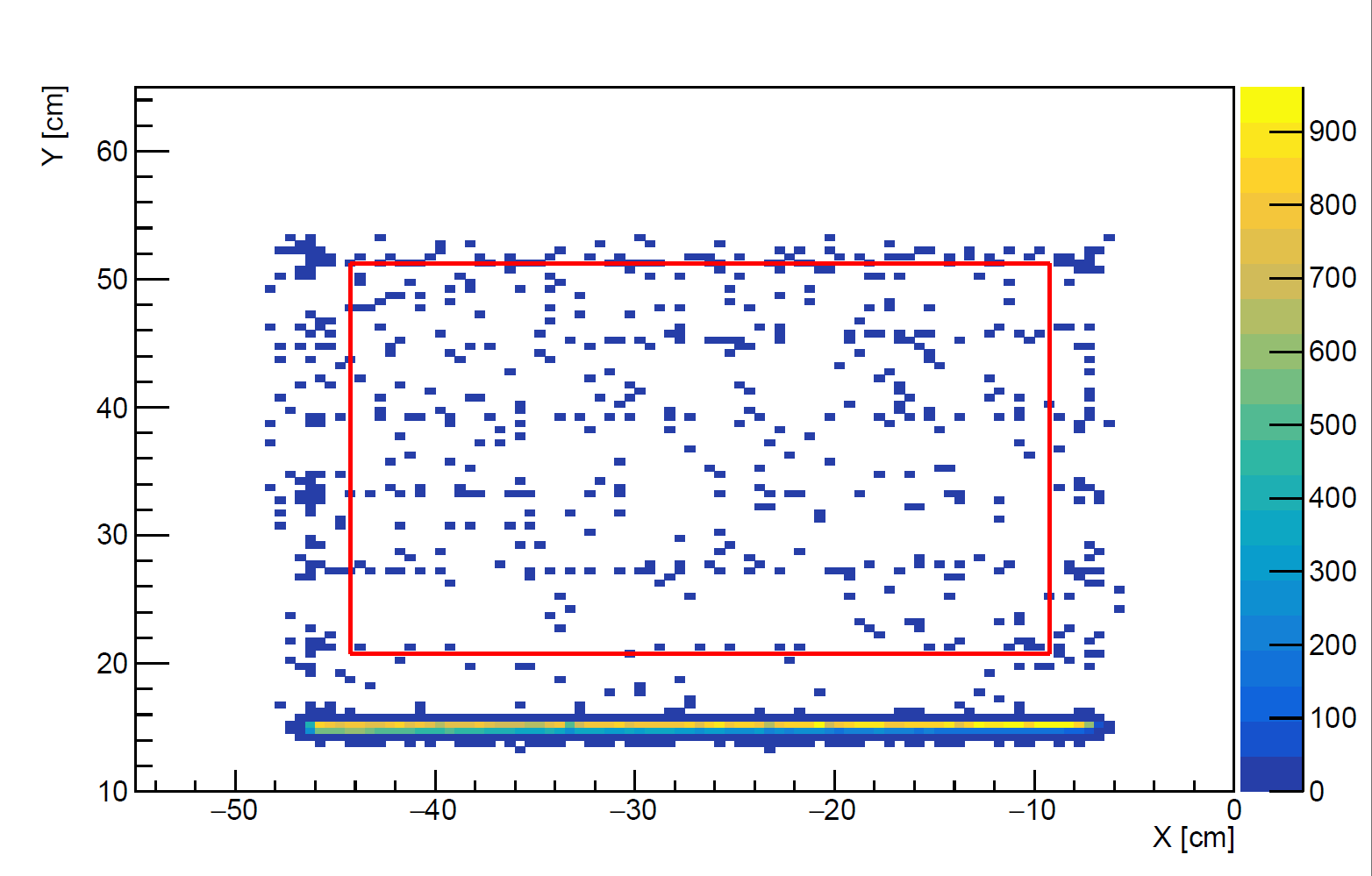}
\caption{The extrapolated position of the reconstructed Scifi track at Veto plane 1 (left) and Veto plane 2 (right). The red square encloses the fiducial area, which was used for the observation of neutrino interactions~\cite{Albanese:031802}. The density of muon tracks at the bottom is higher due to the convolution of the veto
geometrical acceptance and the muon flux distribution measured in the detector. \cite{muonfluxpaper}}
\label{fig:XposVeto12}
\end{figure}

In 2022-2023 the inefficiency of the Veto system was dependent on the data taking condition but the optimal value measured to be $(7.8\pm2.8)\times 10^{-8}$ on a fiducial area of $35\times35~\textrm{cm}^{2}$, corresponding to a target coverage of $\sim 64\%$.
To obtain the required rejection power, the first two SciFi planes were added to the Veto system which significantly reduced the fiducial volume.
The background from muons was now negligible, but this fiducial volume cut rejects 92.4\% of the neutrino charged current interactions. This is because the neutrino density is higher in the bottom part of the detector (see  Figure~\ref{fig:XposVeto12}). 

\begin{figure}[ht] 
\centering
\includegraphics[width=0.7\textwidth]{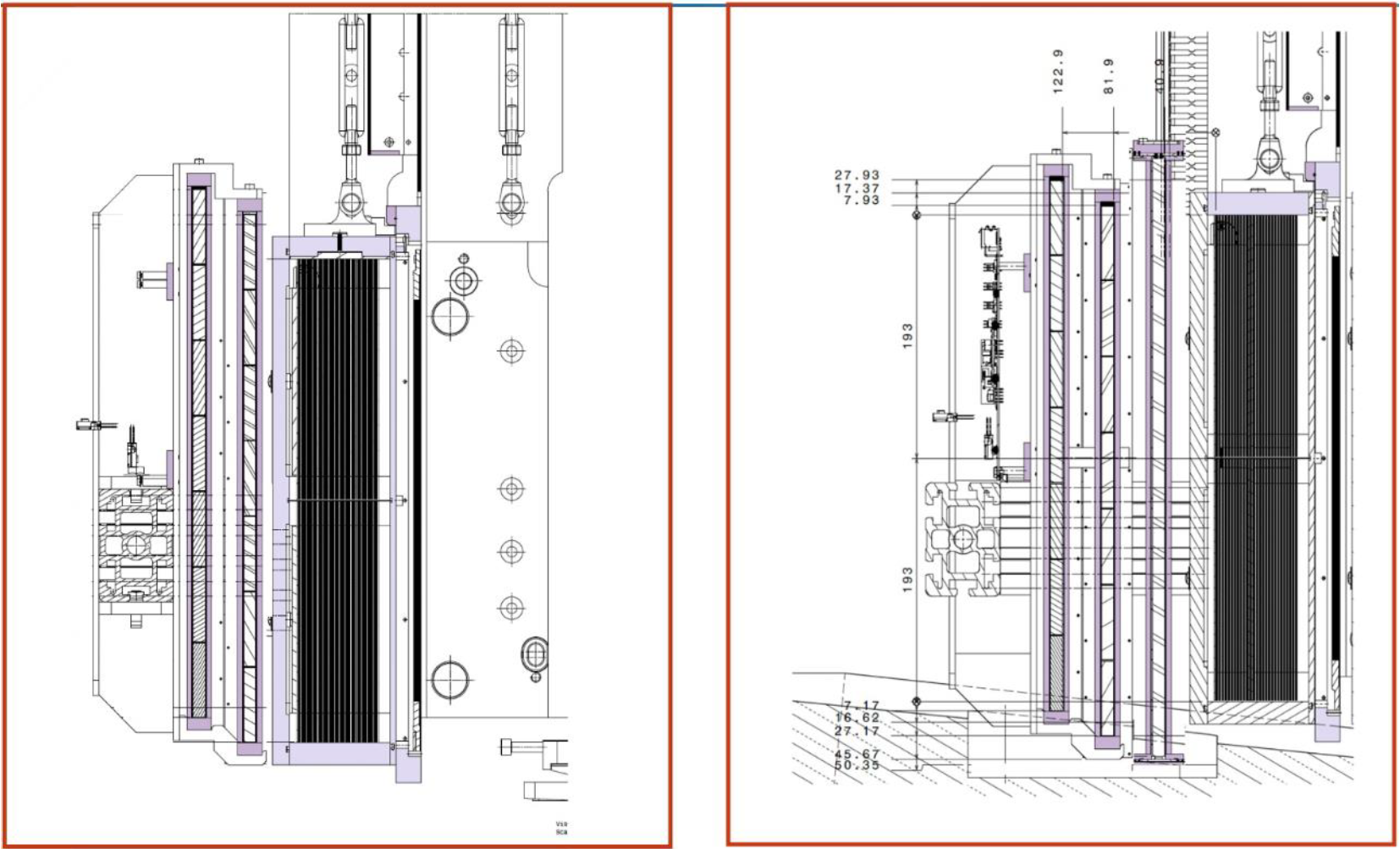}
\caption{Current Veto system layout with two planes with horizontal bars (left). The upgraded Veto system with a third plane with vertical bars (right).}
\label{fig:vetoupgrade}
\end{figure}

To improve the  efficiency of the Veto system, a third Veto station with vertical
bars was installed during the Year End Technical Stop of 2023--20244 (see Figure \ref{fig:vetoupgrade}). Moreover, the whole Veto system was lowered by $\approx{27}\ \rm cm$ to provide full coverage of the target by all stations.

In this paper, we report on the construction of the Veto~3 plane, its performance during commissioning with cosmic rays, its installation in TI18, the measured system inefficiency results, and provide an estimate of the expected increase in the number of observed neutrino interactions.

\section{Design and construction}
\label{sec:design}

The Veto~3 module, made of scintillating bars read out by SiPMs, uses the same technology as Veto~1 and 2~\cite{SNDLHCdetectorpaper}.

\begin{figure}[ht]
\centering
\includegraphics[width=0.8\textwidth]{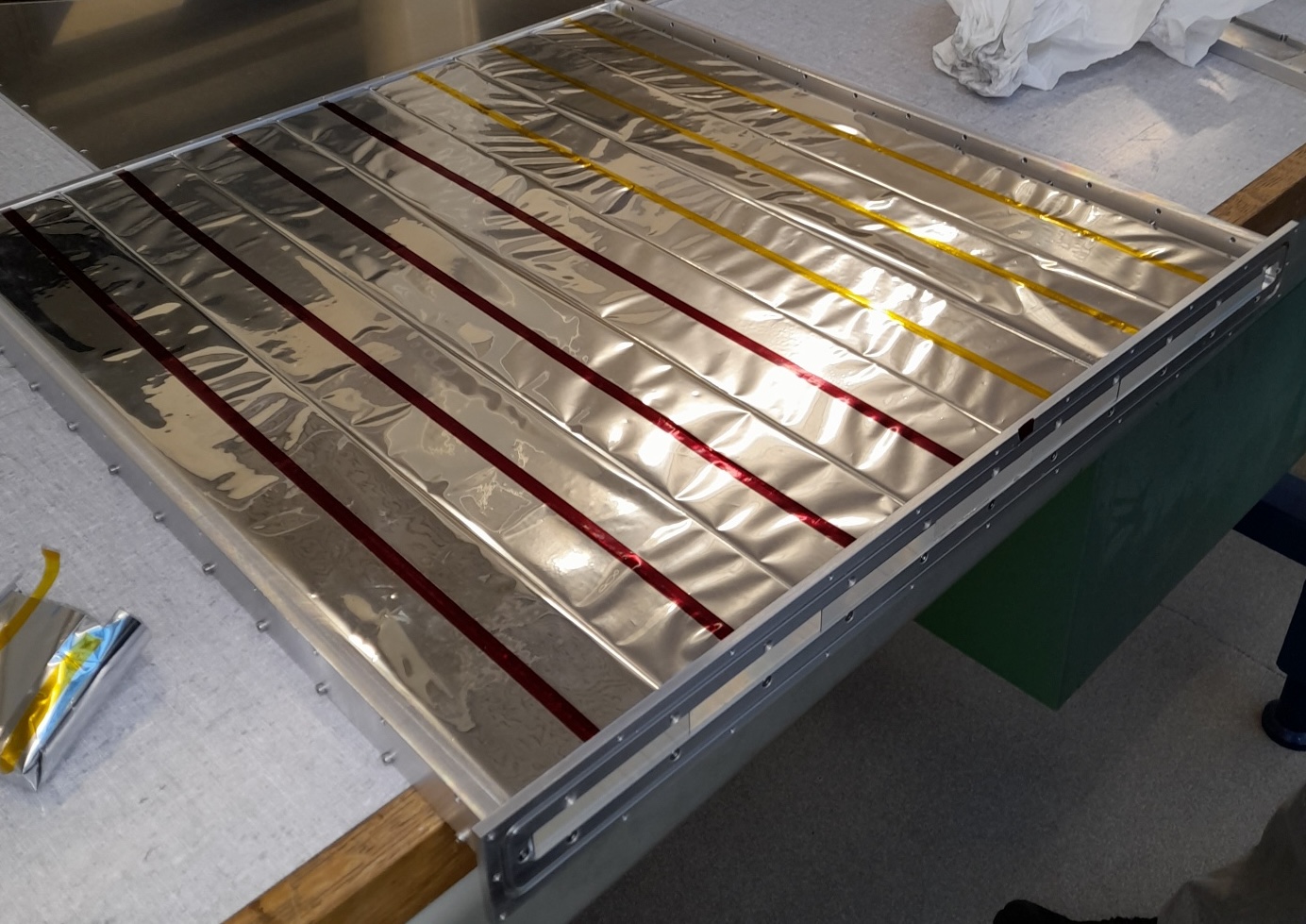}
\includegraphics[width=0.8\textwidth]{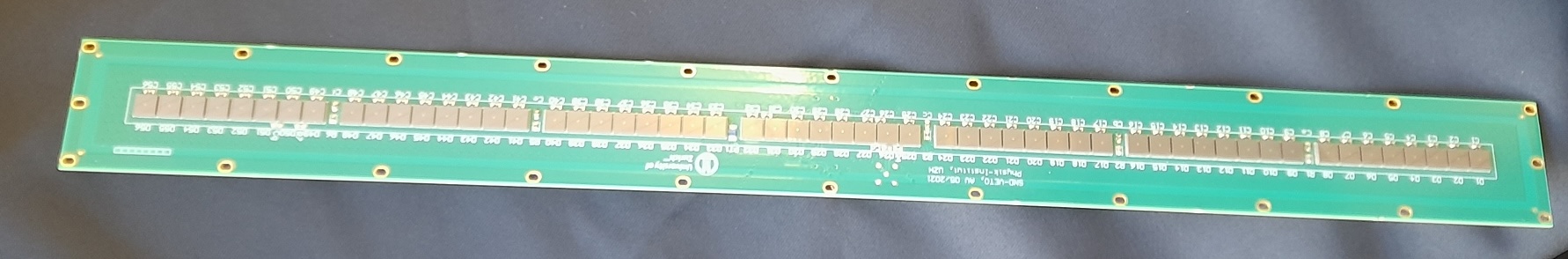}
\caption{ Top: the seven scintillating bars wrapped in aluminized mylar and aligned side by side. Bottom: PCB hosting the $56$ SiPMs.}
\label{fig:barsandRO}
\end{figure}

While Veto~1 and 2 have horizontally disposed bars, the Veto~3 bars are aligned vertically (see Figure~\ref{fig:barsandRO}, top).
The Veto~3 bars are $46\ \rm cm$ long, extending by $2\ \rm cm$ above and below the Veto 1 and 2 range
(see Figure~\ref{fig:vetoupgrade}).

The Veto~3 bars are read out at the top edge by $56$ Hamamatsu Photonics MPPC S14160-6050HS SiPMs ($6\times6 \rm{mm}^2, 50 \mu\rm{m}$ pitch)\cite{sipm_6050_url} 
hosted on a printed circuit board (PCB) and subdivided in seven equally spaced bunches, which are carefully aligned to be centered between bar edges, so that each bar is uniquely read out by 8 SiPMs
(see Figure~\ref{fig:barsandRO}, middle).

To ensure optimal readout the bar edges are placed in close contact with the SiPMs without using optical gel, while the other side is carefully closed with flat aluminized mylar pieces to act as efficient mirrors.

As for Veto 1 and 2, the SiPMs signals are routed to the front-end PCB, which is mounted on the Veto~3 module and hosts two TOFPET ASICs \cite{tofpet}, each viewing $28$ SiPMs.

The vertical disposition of bars in Veto~3 minimises the probability that the readout dead time coincides with the passage of a second muon through the same bar. Although Veto~1 and Veto~2 are staggered, there remains a possibility that two muons could traverse the overlapping region. If the time interval between the two particles is sufficiently short, the signal from the second muon may be lost due to the time required to process the signal from the first. The vertical bars in Veto~3 provide an additional vertical segmentation, thereby reducing the overlapping region among the three bars. This minimizes the probability of two muons hitting the detector so close in space and time that no signal is left in any of the Veto planes.

The completed Veto~3 module was tested for electrical and light tightness. The SiPM dark count rate was measured to set a suitably low noise level.

\section{Performance with cosmic rays }
\label{sec:performance}
The Veto~3 module was first tested with a dedicated laboratory setup to check that all channels were efficient and to look for systematic asymmetries in the efficiency along the bars, between the bar end close to the SiPMs and the opposite reflective end.

The setup that was used for this test consisted of the Veto~3 module and a SND@LHC DownStream (DS) chamber.

The DS chamber consists of $82.5\times1\times1 \,\rm  cm^{3}$ scintillator bars arranged in two layers to cover both $x$ and $y$ views, called horizontal (H) or vertical (V). 
The vertical plane (DS V) is instrumented with one SiPM at the top end of each bar, whereas the horizontal plane has two SiPMs, one at each end (DS R and DS L).
To collect cosmic ray data, the DS was placed parallel to the floor, and the Veto~3 plane was positioned horizontally above the DS chamber. 

A cosmic ray event was tagged by requiring:
\begin{itemize}
    \item[-] signals from DS R, L and V. If there was more than one hit, we selected the one with the higher value of integrated charge (QDC - Charge to Digital Conversion), i.e. the larger light input to the SiPMs;
    \item[-] the signals in DS R and L had to be originated from the same bar;
    \item[-] all signals to be above a predefined QDC threshold, to further cut noise.
\end{itemize}

\begin{figure}[htbp]
\centering
\includegraphics[width=.45\textwidth]{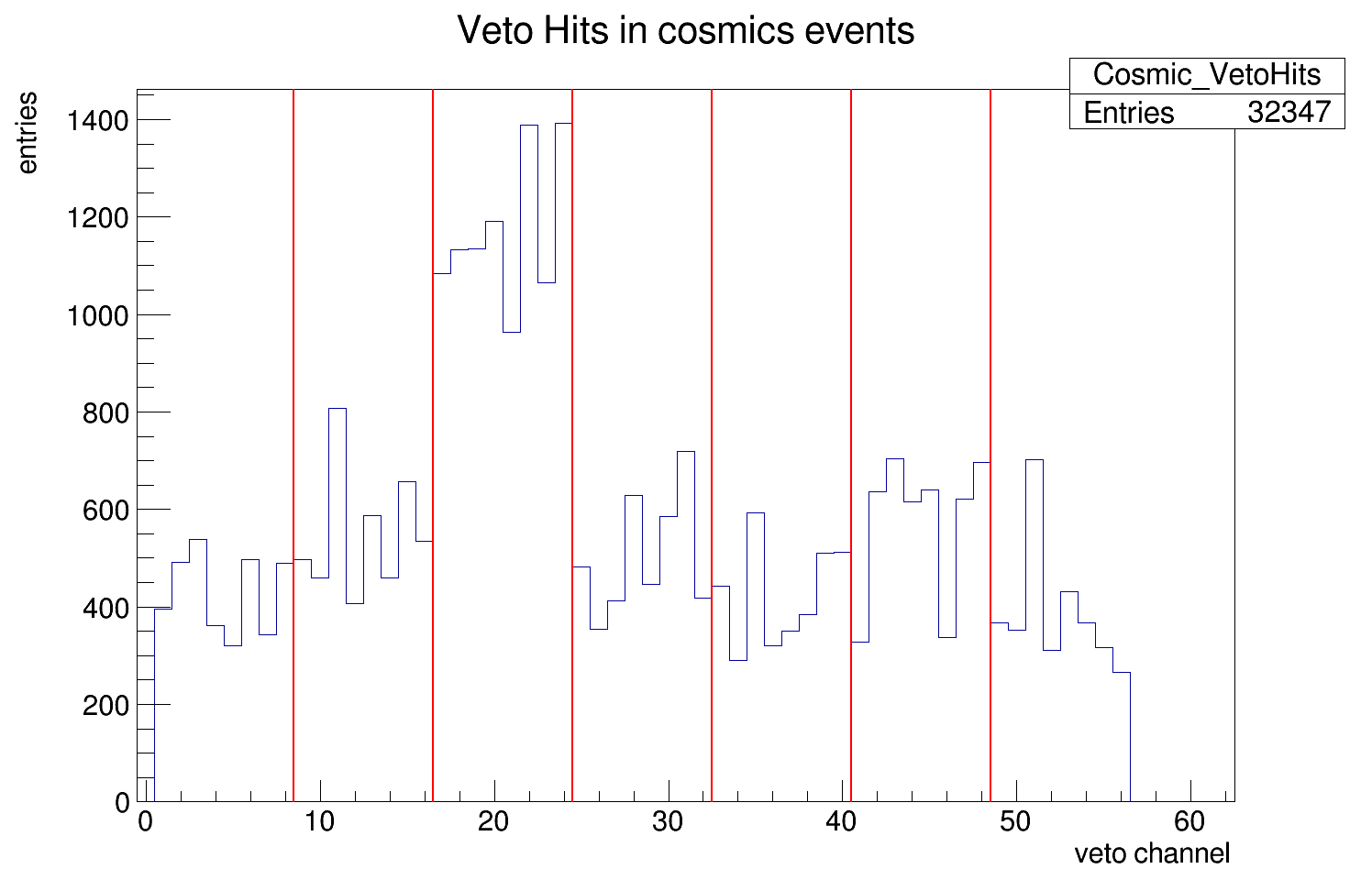}
\includegraphics[width=.45\textwidth]{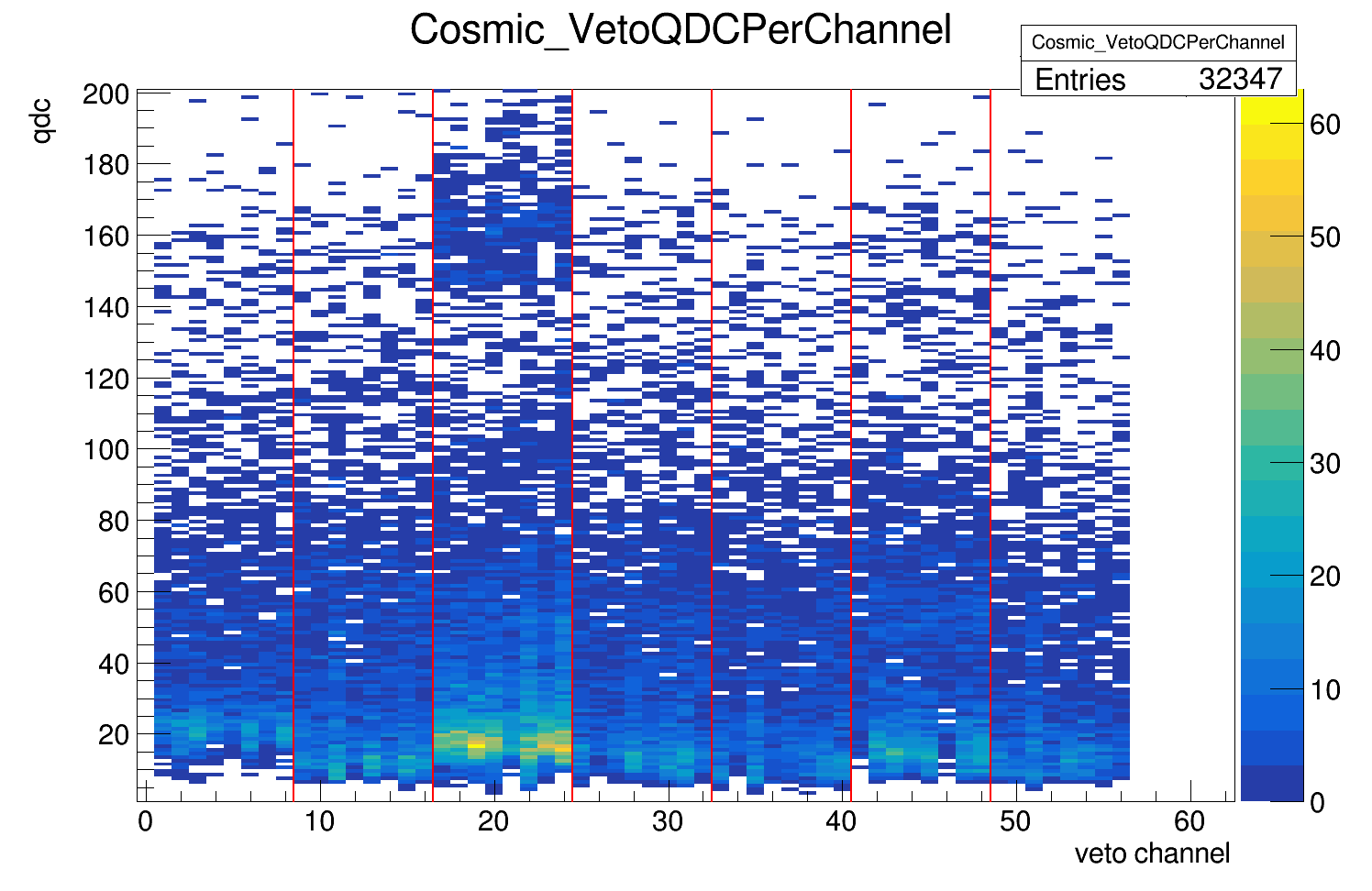}
\caption{\label{fig:blg29} Hit (left) and QDC (right) distribution per Veto channel in the selected subset of cosmic ray events. The red lines delimit the bars.}
\end{figure}

To verify that channels were functional we examined the hit distribution and QDC response. The hit distribution, shown in Figure~\ref{fig:blg29} (left), features a peak in the third bar which is not expected for cosmic rays. It was caused by a gain setting that was too high. This is shown in the QDC distribution per channel, Figure~\ref{fig:blg29} (right), in which a saturation effect is visible for this bar.

To reduce the contribution of background and noise when computing the efficiency, the bar hit in the Veto must be spatially compatible with the expected cosmic ray position measured using the DS chamber. Moreover, a minimum QDC threshold was set for the veto signal. 

To investigate a possible dependence of the efficiency on the position of the hit along the bar caused by the single-sided readout, the hit rate as a function of the hit position was studied. The Veto~3 area was divided into 3 bands of $12 \,\rm cm$, measured from the readout side of the bars. To avoid any effect due to the lower efficiency expected at the borders, the first and last three cm were neglected in the computation.

\begin{figure}[htbp]
\centering
\includegraphics[width=.65\textwidth]{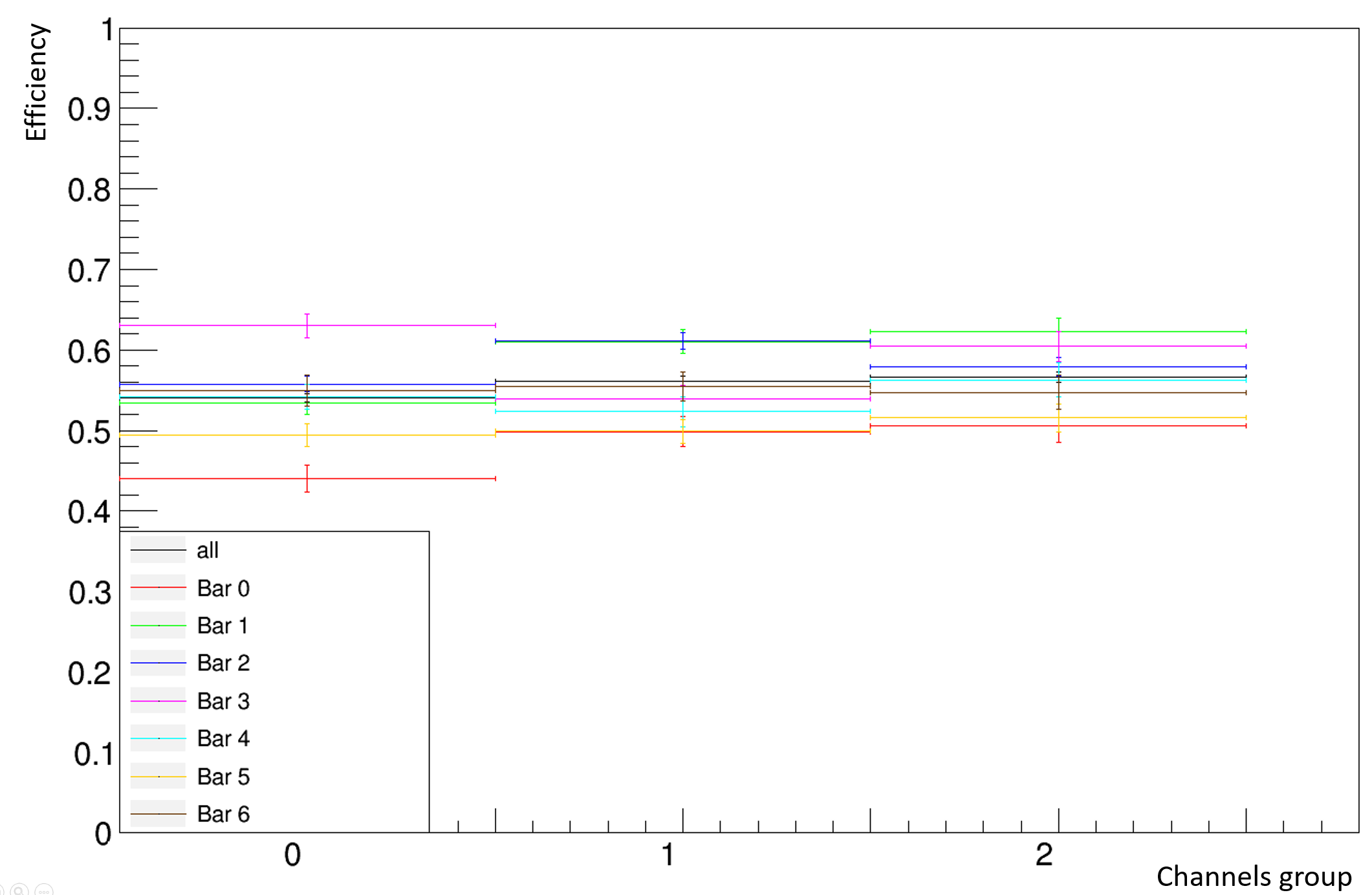}
\caption{\label{fig:blg29_eff} Efficiency Veto~3 bars as a function of the hit position. 
Each bin is the average of $12 \,\rm cm$. Bin 0 is the side furthest from the readout SiPM. Bin 1 is the central region. The last group is closest to the readout side. }
\end{figure}

The plot in Figure~\ref{fig:blg29_eff} shows the efficiency of each bar as well as the overall detector one as a function of the hit position. As expected, there is a minor and negligible effect due to the distance from the readout. Indeed, the relative efficiency
\begin{equation*}
    \rm {relative\, efficiency} = \frac{\rm{efficiency\, far\, from\, SiPM}}{\rm {efficiency\, close\, to\, SiPM}} = 0.95 \pm 0.02
\end{equation*}

No issues were identified in the cosmic rays test in the laboratory, and the Veto~3 plane was cleared for installation in TI18.

\section{Installation and commissioning}
\label{sec:installation}
With the addition of the third plane, the overall thickness of the Veto system increased, requiring an adjustment of its position in both the upstream and vertical directions; to lower the system sufficiently to cover the target (see Figure~\ref{fig:vetoupgrade}), a trench was cut into the concrete floor.

To lower the Veto system to cover the target (see Figure~\ref{fig:vetoupgrade}), a trench had to be cut out in the concrete floor. 

\begin{figure}[htb]
    \centering
    \includegraphics[width=.6\textwidth]{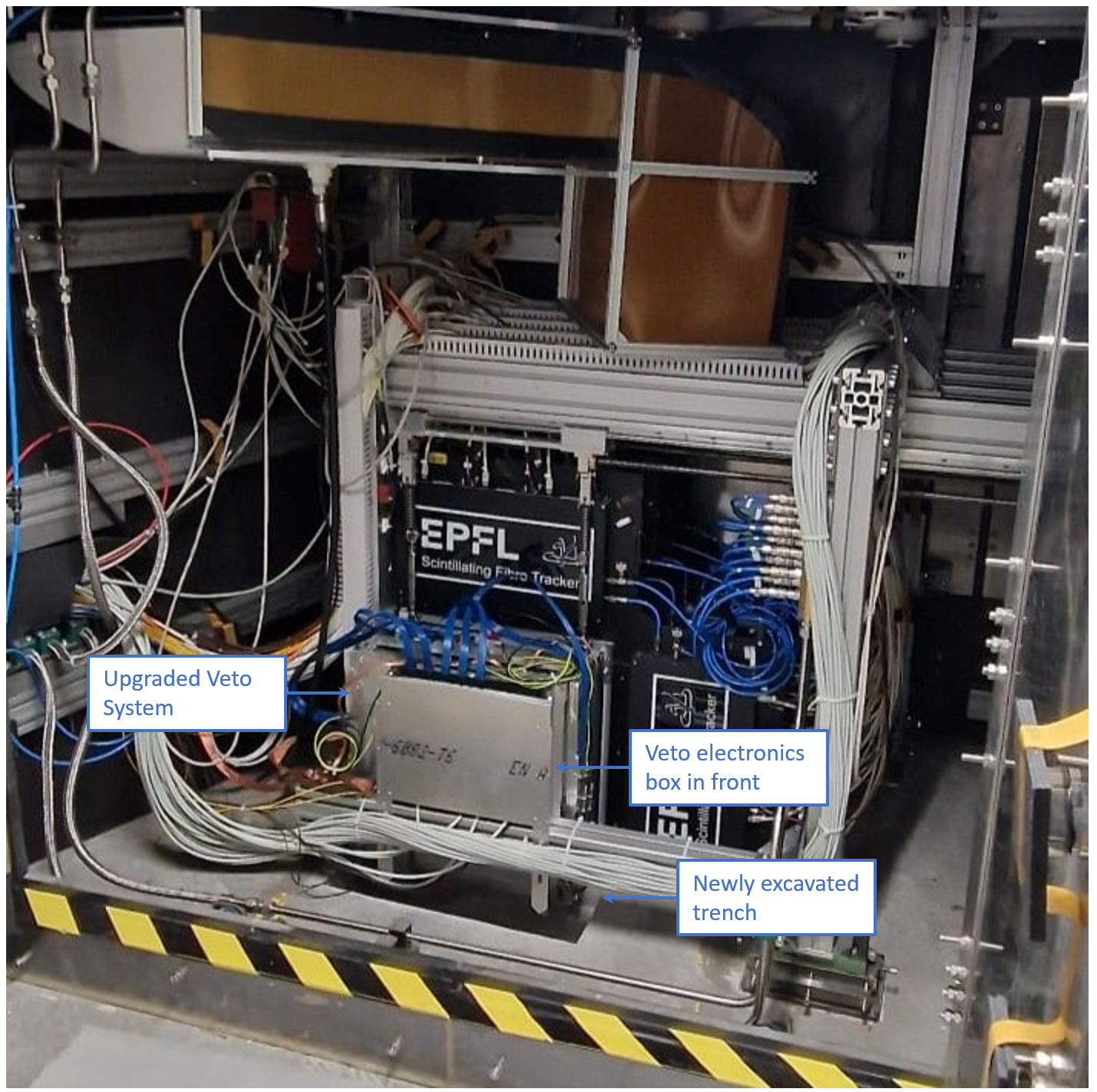}
    \caption{Final configuration of SND@LHC detector with upgraded Veto system.}
    \label{fig:SND_final_view}
\end{figure}

After completing this civil engineering work, the upgraded Veto system was installed. 
The final configuration is shown in Figure~\ref{fig:SND_final_view}, which displays the new trench, the position of the Veto system relative to the rest of the detector, and the box housing its readout electronics.

\begin{figure}[h]
\centering
\includegraphics[width=.45\textwidth]{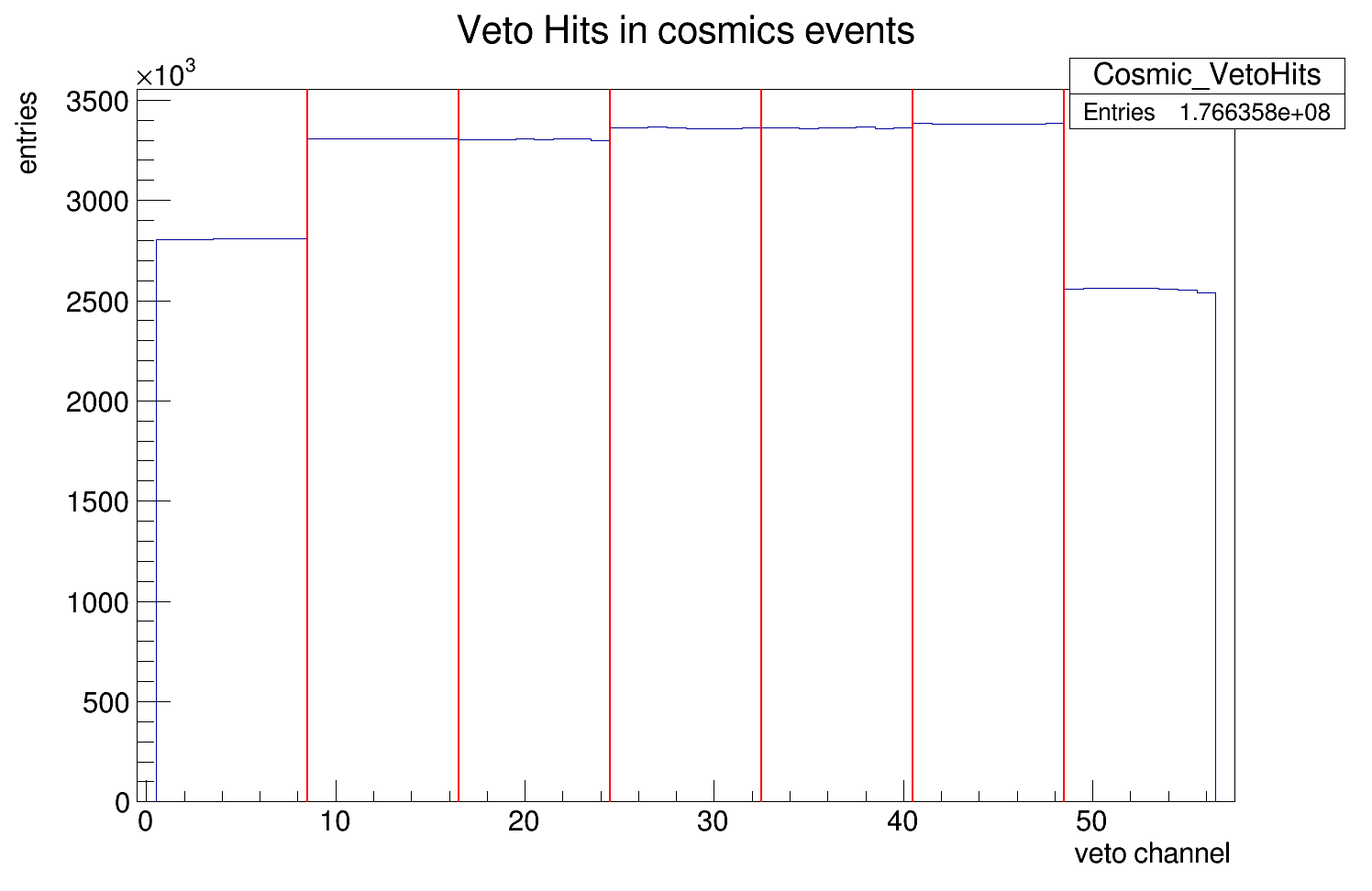}
\includegraphics[width=.45\textwidth]{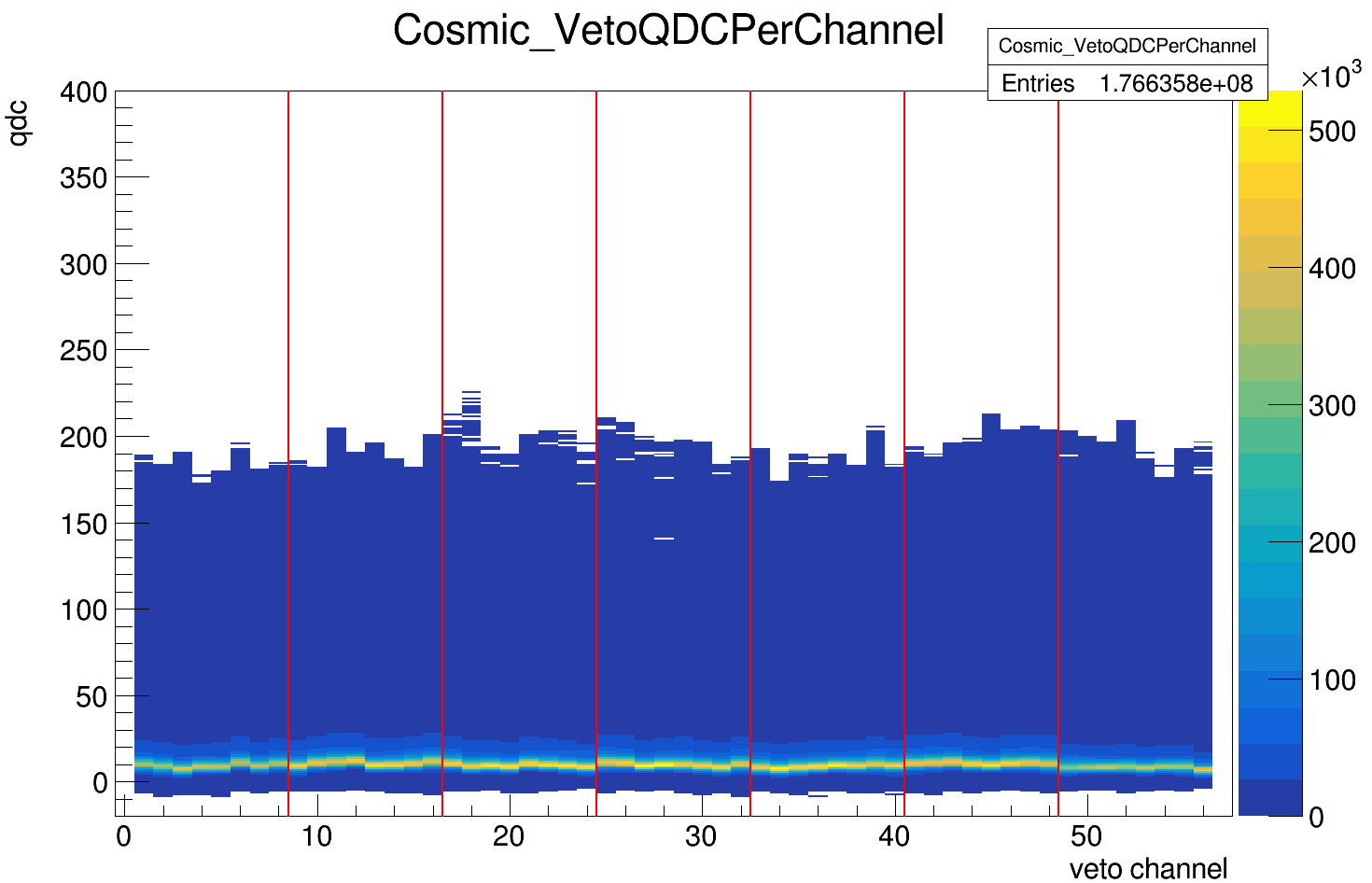}
\caption{Veto~3 Hit (left) and QDC (right) distribution in cosmic ray events selected in TI18. The red lines delimit the scintillator bars.}
\label{fig:ti18_hit_QDC}
\end{figure}

The Veto~3 was commissioned using cosmic rays. The hit distribution per channel (see Figure~\ref{fig:ti18_hit_QDC}) shows that the Veto plane was not damaged during transport and installation and that all channels were correctly taking data.  
The thresholds and gain settings adopted in the TI18 configuration effectively address the issues of electronic noise and QDC saturation that were identified during the laboratory testing phase (see Section~\ref{sec:performance}). By carefully calibrating and tuning these parameters in TI18 conditions, the performance limitations encountered in the initial setup were mitigated, resulting in reduced noise and no QDC saturation.

\begin{figure}[htbp]
\centering
\includegraphics[width=.65\textwidth]{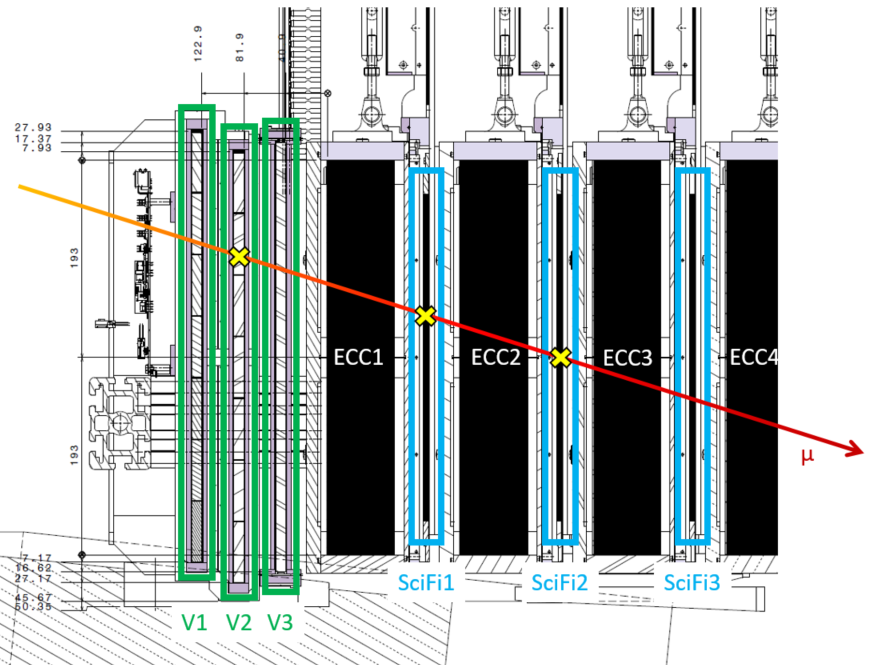}
\caption{Schematic representation of the selection of cosmic ray events. The Veto planes are highlighted in green. The SciFi modules are identified in blue. }
\label{fig:CR}
\end{figure}

Cosmic ray events were selected by requiring signals in Veto~2 and the first two SciFi modules as shown in Figure~\ref{fig:CR}.
This allowed the computation of the expected signal position, ensuring that the particle trajectory crossed the Veto plane under study. 
An additional requirement on the maximum incoming angle of the muon was introduced to avoid a bias in the efficiency calculation.  In particular, we required a maximum displacement of $1.5 \,\rm cm$ in the $x$ and $y$ directions between the hits in the two SciFi planes.
These requirements are the same for detecting muons from LHC collisions since they traverse the detector horizontally. 
The Veto plane was considered efficient if there was a hit and the signal was found in the expected bar or the neighbouring ones. 

\begin{figure}[htbp]
\centering
\includegraphics[width=.65\textwidth]{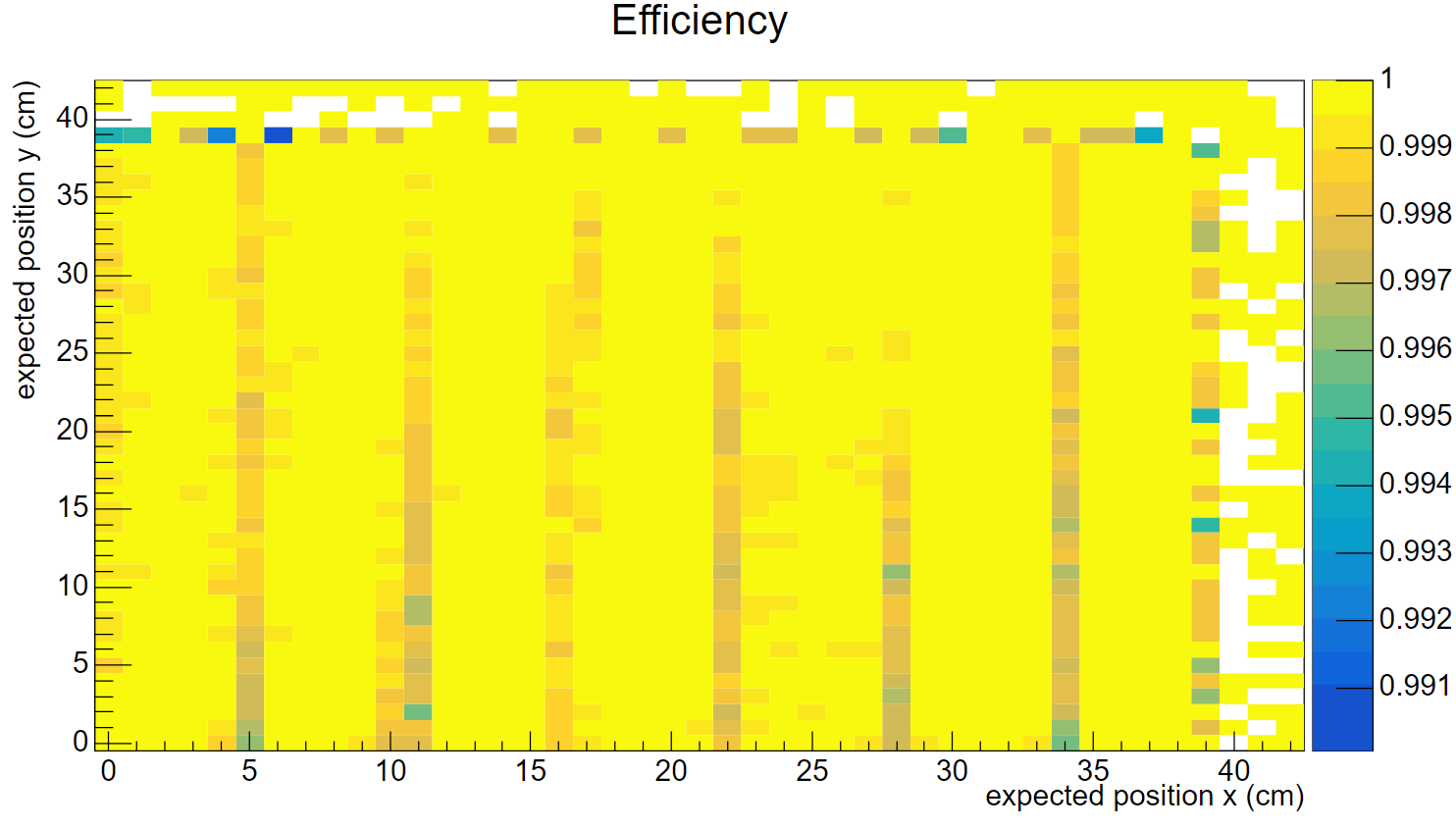}
\caption{Veto~3 efficiency measured during commissioning.}
\label{fig:Ti18_eff}
\end{figure}

The efficiency distribution is shown in Figure \ref{fig:Ti18_eff}. It is uniform and close to 1 in all detector areas. The vertical drops correspond to the bar edges as expected, while the lines at $x \simeq 38 \, \rm cm$ and $y \simeq 38 \, \rm cm$ delimit the SciFi acceptance.
This also confirms that the single-side readout does not affect the efficiency, as already indicated by the preliminary result given in Section~\ref{sec:performance}.
\section{Measured Veto inefficiency }
\label{sec:measuredineff}
The individual Veto plane inefficiencies are estimated using passing through muons by extrapolating reconstructed tracks in the SciFi and the DS to each plane. A plane is considered inefficient if it registers no hit where one is expected because of a passing track (see~\cite{notaveto3}).

The results shown here are obtained using $102 \,\rm{fb}^{-1}$ of the delivered luminosity of which $8\times10^{8}$ high-quality passing through muon events were selected.

\begin{figure}[htbp]
\centering
\includegraphics[width=0.32\textwidth]{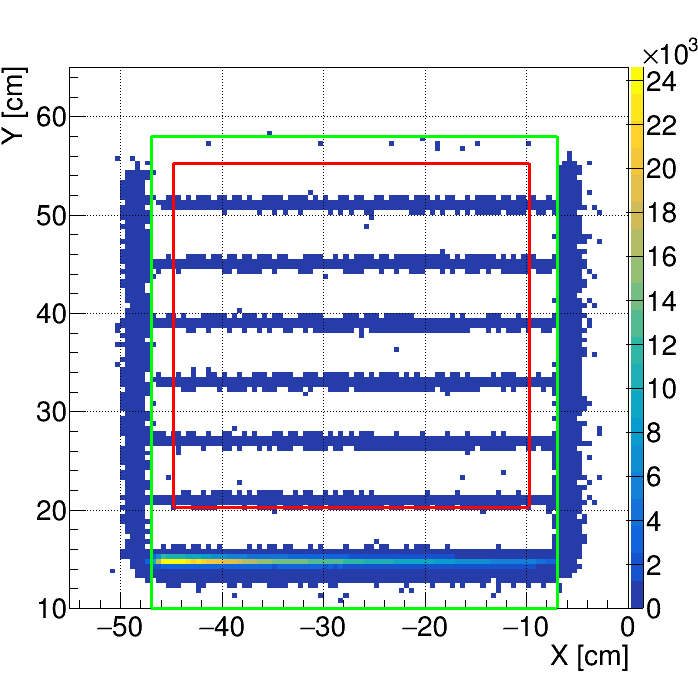}
\includegraphics[width=0.32\textwidth]{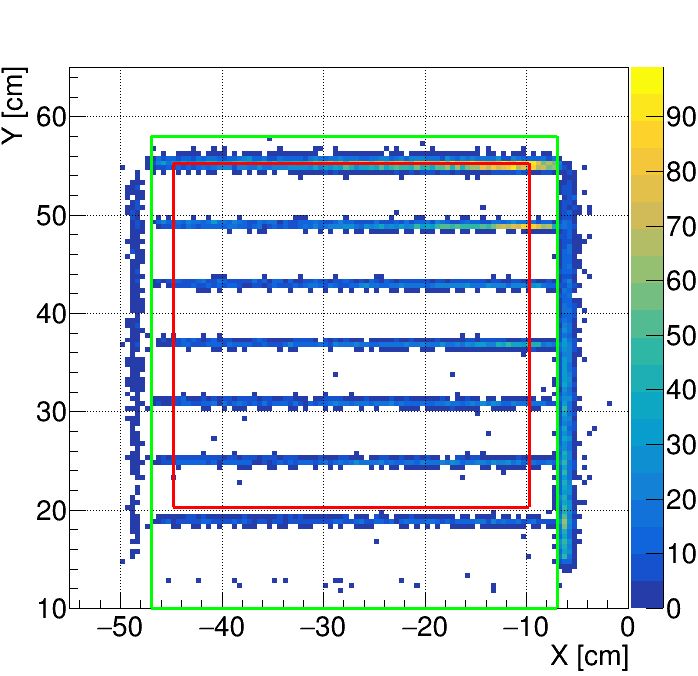}
\includegraphics[width=0.32\textwidth]{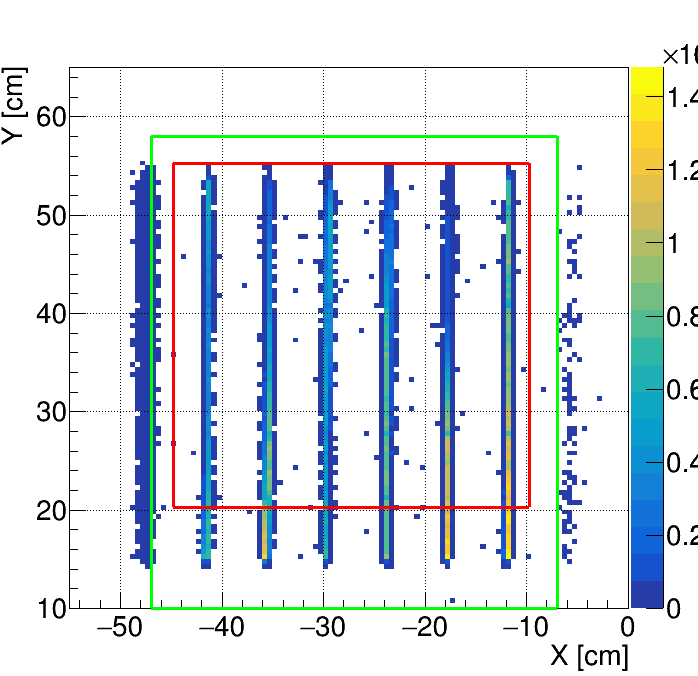}
\caption{The extrapolated position of the reconstructed SciFi tracks at Veto~1 (left) and Veto~2 (middle) and Veto~3 (right) for events without any fired Veto channel in that plane. The red square encloses the fiducial area used for the 2022-2023 analyses; the green rectangle delimits the fiducial area used in 2024.}
\label{fig:veto_plane_ineff}
\end{figure}

\begin{figure}[htbp]
\centering
\includegraphics[width=.35\textwidth]{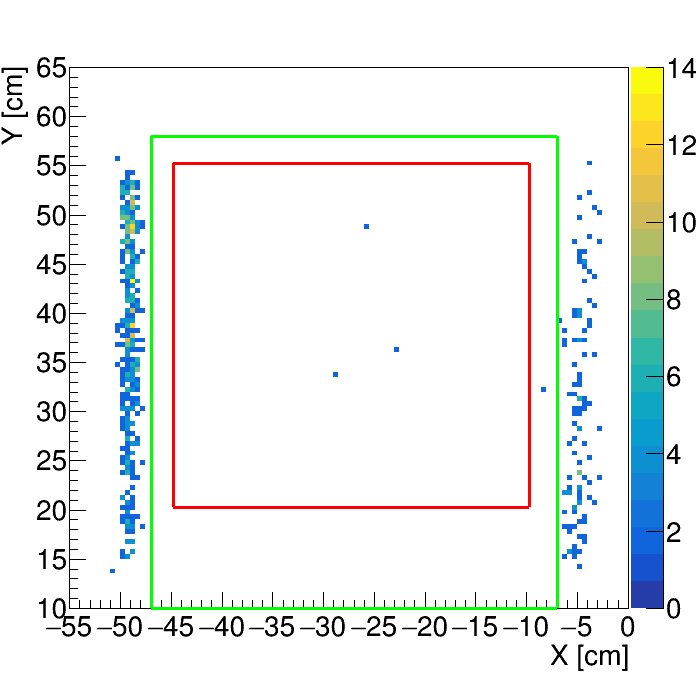}
\caption{The extrapolated position of the reconstructed SciFi tracks at Veto~1 for events without any fired Veto channel in all three planes. The red square encloses the fiducial area used for the 2022-2023 analyses, while the green rectangle represents the fiducial area used in the 2024 Veto inefficiency study.  }
\label{fig:veto_system_ineff}
\end{figure}

The x-y positions of SciFi tracks for which there is no corresponding Veto plane hit are shown in Figure~\ref{fig:veto_plane_ineff}. The dead areas in between scintillating bars are visible. As for 2022 and 2023 (see Figure~\ref{fig:XposVeto12}) there are areas outside the acceptance of a given Veto plane which are densely populated with tracks, especially the horizontal band below Veto~1. However, these regions are enclosed in the sensitive areas of the other planes, as shown in the combined system plot in Figure~\ref{fig:veto_system_ineff}. The selected area of best system performance, a fiducial area of $40\times48~\textrm{cm}^{2}$ in ranges $-47~\textrm{cm}\leq\textrm{x}\leq-7~\textrm{cm}$ and $10~\textrm{cm}\leq\textrm{y}\leq58~\textrm{cm}$, shown as green rectangles in Figures~\ref{fig:veto_plane_ineff}~and~\ref{fig:veto_system_ineff}  yields an overall Veto system inefficiency of $(4.9\pm1.9)\times10^{-9}$. 

\begin{figure}[htbp]
\centering
\includegraphics[width=.55\textwidth]{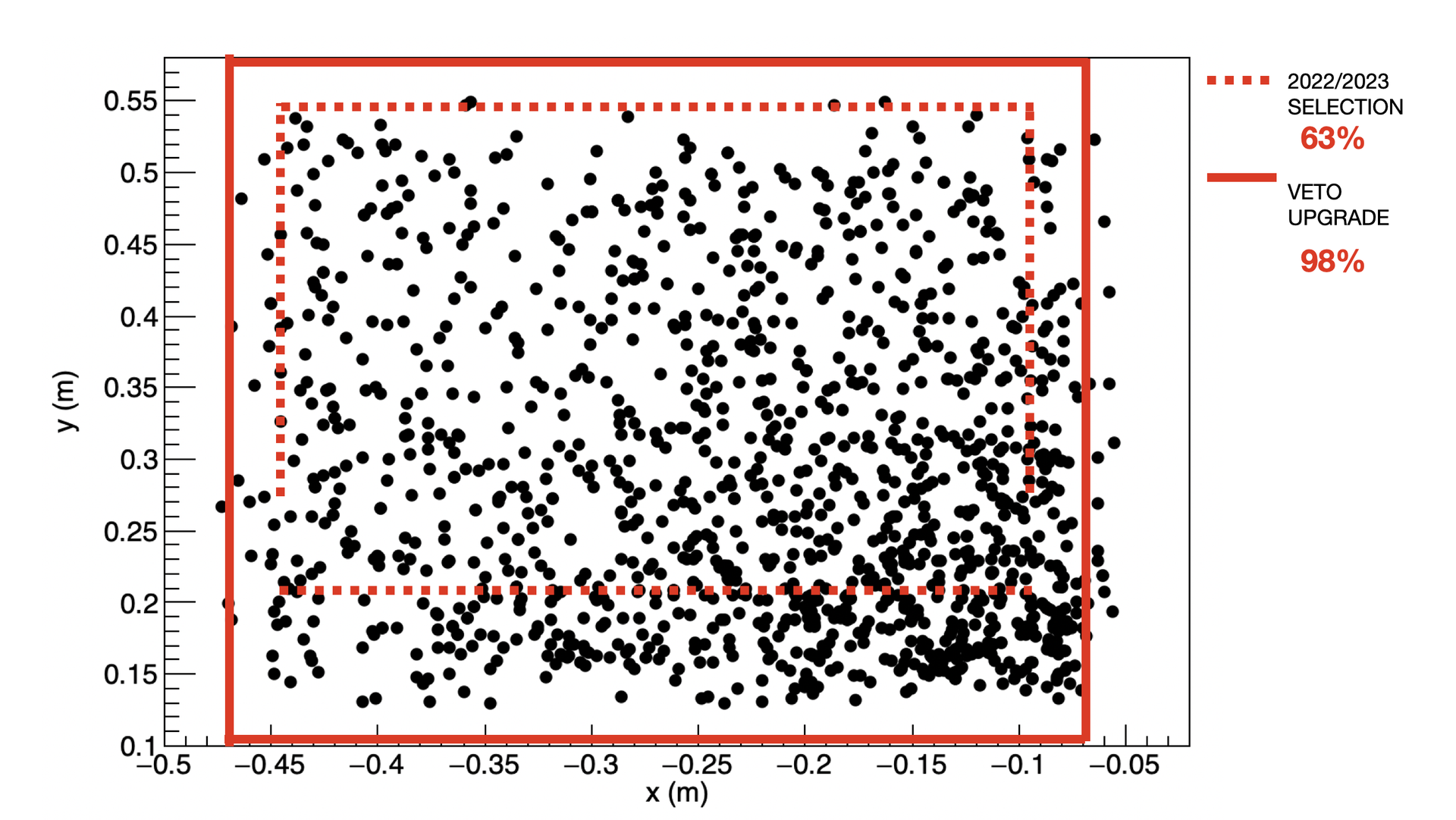}
\caption{Position of muon neutrino interactions from Monte Carlo simulation. The dashed line represents the fiducial area used in 2022-2023, while the solid one encloses the acceptance of the upgraded Veto system.  }
\label{fig:neutriniMC}
\end{figure}

The fiducial area for neutrino interactions is now $1.6$~times larger than the area used in the 2022-2023 analysis and fully covers the target region.  Monte Carlo simulation studies suggest that this could lead to an increase of $56\%$ of observable neutrino interactions, see Figure~\ref{fig:neutriniMC}.

\newpage

\section{Summary }
\label{sec:conclusions}

In this paper, we reported the construction of the Veto~3 and its performance during commissioning, and the relocation of the Veto system in TI18 during the 2023-2024 YETS.

The upgraded Veto system fully covers the target region and allows for extending the fiducial acceptance to neutrino interactions into the whole target volume. 

After lowering the Veto system and adding a third Veto module with vertical bars, the optimal inefficiency of the Veto system was reduced from $(7.8\pm2.8)\times10^{-8}$ in 2022-2023 on a fiducial area corresponding to $64\%$ of the target to $(4.9\pm1.9)\times10^{-9}$ in 2024 with full target coverage.
Monte Carlo simulation studies suggest that the larger fiducial volume and reduced inefficiency could give an increase of $56\%$ of observable neutrinos.

\acknowledgments

We acknowledge the support for the construction and operation of the SND@LHC detector provided by the following funding agencies:  CERN;  the Bulgarian Ministry of Education and Science within the National
Roadmap for Research Infrastructures 2020–2027 (object CERN); ANID—Millennium Program—$\rm{ICN}2019\_044$ (Chile); J.C.~Helo~Herrera  and O.~Soto~Sandoval  acknowledge support from ANID FONDECYT grant No.1241685; the Deutsche Forschungsgemeinschaft (DFG, ID 496466340); the Italian National Institute for Nuclear Physics (INFN); JSPS, MEXT, the~Global COE program of Nagoya University, the~Promotion
and Mutual Aid Corporation for Private Schools of Japan for Japan;
the National Research Foundation of Korea with grant numbers 2021R1A2C2011003, 2020R1A2C1099546, 2021R1F1A1061717, and~
2022R1A2C100505; Fundação para a Ciência e a Tecnologia, FCT (Portugal), 
CERN/FIS-INS/0028/2021; the Swiss National Science Foundation (SNSF); TENMAK for Turkey (Grant No. 2022TENMAK(CERN) A5.H3.F2-1).
M.~Climesu, H.~Lacker and R.~Wanke are funded by the Deutsche Forschungsgemeinschaft (DFG, German Research Foundation), Project 496466340. We acknowledge the funding of individuals by Fundação para a Ciência e a Tecnologia, FCT (Portugal) with grant numbers  CEECIND/01334/2018, 
CEECINST/00032/2021 and PRT/BD/153351/2021.

We express our gratitude to our colleagues in the CERN accelerator departments for the excellent performance of the LHC. We thank the technical and administrative staff at CERN and at other SND@LHC institutes for their contributions to the success of the SND@LHC efforts. We thank Luis Lopes, Jakob Paul Schmidt and Maik Daniels for their help during the~construction.


\bibliographystyle{JHEP}
\bibliography{biblio.bib}


\end{document}